\documentclass [aps,pra,amssymb,amsmath,twocolumn,showpacs]{revtex4-1}

\usepackage{amsmath}
\usepackage{amssymb}
\usepackage{graphicx}
\usepackage{verbatim}
\usepackage{bm}
\usepackage[dvipsnames]{xcolor}

\usepackage{graphicx,epsfig,amsfonts,amssymb}
\usepackage{hyperref}
  \hypersetup{colorlinks=true,citecolor=blue}

\usepackage{bm}
\usepackage{times}
\usepackage{lipsum}

\usepackage[all]{xy}

\newcommand{\la}{\langle}
\newcommand{\ra}{\rangle}
\newcommand{\rar}{\rightarrow}

\newcommand{\ben}{\begin{eqnarray}}
\newcommand{\een}{\end{eqnarray}}

\newcommand{\be}{\begin{equation}}
\newcommand{\ee}{\end{equation}}

\begin{document}

\title{Integrability in the multistate Landau-Zener model with time-quadratic commuting operators}
\author{Vladimir Y. Chernyak}
\affiliation{Department of Chemistry, Wayne State University, 5101 Cass Ave, Detroit, Michigan 48202, USA}
\affiliation{Department of Mathematics, Wayne State University, 656 W. Kirby, Detroit, Michigan 48202, USA}
\author{Nikolai A. Sinitsyn}
\affiliation{Theoretical Division, Los Alamos National Laboratory, Los Alamos, New Mexico 87545, USA}

\date{\today}

\begin{abstract}
 All currently known exactly solvable  multistate Landau-Zener (MLZ) models are associated with families of operators that commute with the MLZ Hamiltonians and depend on time  linearly.
There can also be operators that satisfy the integrability conditions with the MLZ Hamiltonians but depend on time quadratically. We show that, among the MLZ systems, such time-quadratic operators are much more common. We demonstrate then that  such  operators generally lead  to  constraints  on the independent variables that parametrize the scattering matrix. Such constraints lead to asymptotically exact expressions for the transition probabilities in the adiabatic limit of a three-level MLZ model. New more complex fully solvable MLZ systems are also found. 

\end{abstract}

\maketitle

\section{Introduction}
 In quantum mechanics, the evolution with a time-dependent Hamiltonian $H(t)$ is described by the nonstationary Schr\"odinger equation
 \be
i\frac{d}{dt}|\Psi (t) \ra =H(t) |\Psi (t)\ra.
 \label{nst-se}
 \ee
 A nonperturbative and nonadiabatic time-dependence of the Hamiltonian parameters in (\ref{nst-se})  destroys the energy conservation and  often leads to fast entanglement generation. This strongly restricts applicability of numerically approximate techniques such as tensor networks and DFT. A common intuition about stationary quantum mechanics usually, for such systems, does not apply. For example, an 
  operator ${I}(t)$ that commutes with a time-dependent Hamiltonian ${H}(t)$, for all times $t$, does not describe a conserved quantity anymore. For solving the corresponding time-dependent Schr\"odinger equation,  knowledge of the eigenvalues and the eigenstates of ${H}(t)$ is usually useless.
  This is why a multistate evolution with  explicitly time-dependent parameters is  hard to study both analytically and numerically.

Nevertheless,  a considerable progress can be achieved in studies of time-dependent Hamiltonians  that belong to the families of the  Hermitian operators $H_j$ with   conditions \cite{commute}
\begin{eqnarray}
 \label{cond1}
 \frac{\partial H_i}{\partial \tau_j} - \frac{\partial H_j}{\partial \tau_i} &=&0,  \quad i,j =1, \ldots, M \ge 2,\\
 \label{cond2}
 [H_i,H_j]&=&0,
\end{eqnarray}
where the dependence of $H_j$ on its parameters is known analytically, and $\tau_j$ is the  parameter that can be treated as the true physical time for the time-evolution with the Hamiltonian $H_j$.

Integrability conditions (\ref{cond1}),~(\ref{cond2}) have been used broadly in the theory of solitons \cite{faddeev-book}. More recently, they were used  as a starting point for deriving fully solvable quantum models with a simple parameter time-dependence. Thus, a large family of such  models was discovered within the multistate Landau-Zener (MLZ) class of the Hamiltonians \cite{quest-LZ}:
\begin{equation}
 H(t)=Bt+A,
    \label{mlz}
\end{equation}
where $B$ and $A$ are Hermitian  $N\times N$ matrices. $B$ can be always diagonalized in a time-independent basis, which is called {\it diabatic basis}. In what follows, we will assume that all matrices are written in this basis. The diagonal elements of $H(t)$ are then called {\it diabatic energy levels}. They are different from the adiabatic energies, which are the eigenstates of the whole Hamiltonian $H(t)$. For MLZ models, diabatic and adiabatic energies merge with each other asymptotically as $t\rar \pm \infty$.

A model (\ref{mlz}) is called {\it solvable} if for any initial state at $t=-\infty$ we can find an analytical expression for  this state at $t= +\infty$.  For simplicity, here we will be interested only in the {\it transition probabilities}, $P_{m\rar n}$, of that the system is found in any diabatic state $n$ at $t=+\infty$ if at $t= -\infty$ the system was in  diabatic state $m$.

Physically, the MLZ models appear frequently when quantum systems are driven by time-dependent fields with a high amplitude of the field modulation, for example, during measurements of magnetic hysteresis of interacting spin clusters \cite{werns,nanomag} or performing the Landau-St\"uckelberg interferometry of electrons in coupled quantum dots \cite{LZ-interf,peta-mlz,fai-LZ}. Although the amplitude of the field modulation is always finite and the time-dependence of parameters is generally nonlinear, the nontrivial dynamics happens in such experiments, as well as in the MLZ models, essentially during finite time intervals that correspond to a passage through a region in energy space with avoided crossings. During this time, the time-dependence of the physical parameters can be linearized, so that the most important dynamics is well approximated by the time-linear Hamiltonian (\ref{mlz}).

It was noticed in \cite{patra-LZ} that several solvable MLZ models had nontrivial linear or quadratic in time commuting operators.  The authors of \cite{patra-LZ} speculated that this fact could be related to solvability of MLZ models. Indeed, it was explained later \cite{parallel-LZ} why fully solvable MLZ systems should generally have a commuting, with the MLZ Hamiltonian, operator that depends on  time $t$ linearly. The latter operator must also  have a simple dependence on a second time-like variable, $\tau$, 
on which the Hamiltonian in (\ref{mlz}) depends linearly too. Thus, for the MLZ model's complete solvability, it is sufficient that the commuting operator be \cite{parallel-LZ}
\begin{equation}
H'_{t/\tau}(t,\tau) = B_0t +B_1 \tau + A +C/\tau,
    \label{linearH}
\end{equation}
where $B_0$, $B_1$, $A$ and $C$ are real symmetric $(t,\tau)$-independent matrices, such that the Hamiltonians (\ref{mlz}) and (\ref{linearH}) satisfy the integrability conditions (\ref{cond1}) and (\ref{cond2}) for  two time-like variables, $\tau_1=t$ and $\tau_2=\tau$. Even though there are exceptional solvable MLZ models that do not have a commuting operator of the form (\ref{linearH}), all such models could be obtained so far from the models that had the commuting operator (\ref{linearH}) by considering special limits of the parameters.

If the reader is not familiar with the developments on solvable MLZ systems, we advise to study-look in our prior publications \cite{quest-LZ,parallel-LZ}. We also refer to \cite{laser-LZ,dzyarmaga,bcs,gamma-LZ,yuzbashyan-LZ} for applications and other recent developments on this topic.
According to \cite{parallel-LZ},  if one of the Hamiltonians has the MLZ form 
\be
H(t,\tau) =  B(\tau)t+A(\tau),
\label{mlz2}
\ee
where $\tau$ is an arbitrary parameter, then conditions (\ref{cond1})-(\ref{cond2}) can be generally satisfied by a $t$-quadratic polynomial operator 
\begin{equation}
H'(t,\tau)= \frac{\partial_{\tau} B(\tau) t^2}{2} + \partial_{\tau}A(\tau) t+D(\tau),
\label{quadH1}
\end{equation}
where the Hermitian matrices  $B(\tau)$, $A(\tau)$, and $D(\tau)$ have to satisfy the conditions
\ben
\label{ccond1}
\frac{1}{2} [\partial_{\tau} B(\tau),A(\tau)] + [\partial_{\tau}A(\tau),B(\tau)]=0,\\
\label{ccond2}
 [\partial_{\tau} A(\tau),A(\tau)] +[D(\tau), B(\tau)] = 0,\\
 \label{ccond3}
 [A(\tau),D(\tau)]=0,
 \een
and $B(\tau)$ is diagonal.
Equations~(\ref{ccond1})-(\ref{ccond3}) are the most general integrability conditions for all MLZ models. 
Given that the known fully solvable MLZ models are associated with much simpler commuting operators (\ref{linearH}), the notion of integrability is looking more general than the complete model's solvability, i.e.,  these two concepts are not equivalent.

Thus, we formulate the  main questions for the present article: 

(i) how common are the $t$-quadratic commuting operators within the MLZ theory? and, 

(ii) what is generally the advantage of knowing that an explicitly time-dependent quantum model is integrable?

\begin{figure}[t!]
    \centering
    \includegraphics[width=0.45\textwidth]{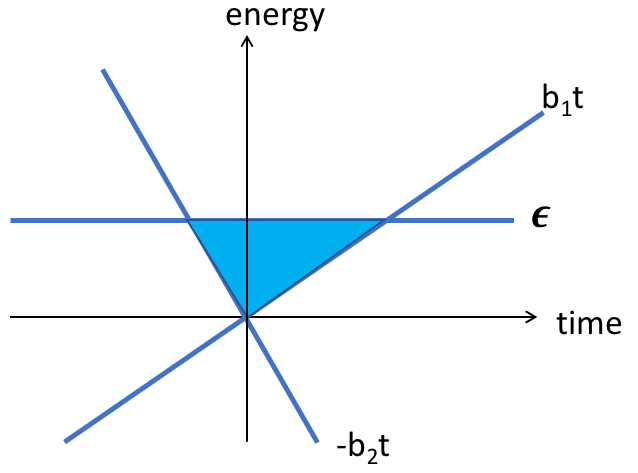}
    \caption{Blue lines are the time-dependent diabatic energy levels of the Hamiltonian (\ref{threeH1}). The blue triangle has these levels as its boundary. The area of this triangle is $S(\tau)=\epsilon(\tau)^2[1/b_1(\tau)+1/b_2(\tau)]/2$. }
    \label{three-levels-pic}%
\end{figure}

We will explore these questions  for the MLZ Hamiltonians ({\ref{mlz2}) by finding $t$-quadratic commuting operators. Having their explicit forms, we will then judge how useful they are for analytical studies of an explicitly time-dependent evolution.
In section~\ref{three-sec}, we will work out the simplest but nontrivial case of the general three-state MLZ model. We will show that this case is integrable but generally not completely solvable. Instead, the integrability leads to independence of the transition probabilities of 
some parameter combination.  We will derive this invariance, discuss its physical meaning, and show how it can be used to reduce the order of the corresponding differential equation. Hence, we will claim that the integrability should be generally understood as a nontrivial 
dynamic symmetry that simplifies the analysis but may or may not lead to the complete model's solution.
In section~\ref{adiab-sec}, we will present the first application of the integrability with $t$-quadratic commuting operators, i.e., to designing the analytical semiclassical approach that leads to asymptotically exact expressions for the transition probabilities in  the MLZ systems in the adiabatic limit. In section~\ref{bipartite-sec}, we demonstrate that the  commuting operators can be found generally in the class of MLZ models with bipartite interactions. In particular, we show the consequences of this integrability in  MLZ chain models with arbitrary number of interacting states, and derive two new fully solvable MLZ models as examples of application of this integrability.    

\section{General three-state Landau-Zener (LZ) model}
\label{three-sec}
As the simplest nontrivial example, we consider the general case of the three-state model with  the  Hamiltonian
\be
H_3=\left( \begin{array}{ccc}
b_1t & g_{12} & g_{13} \\
g_{12}^* & \epsilon & g_{23} \\
g_{13}^* & g_{23}^* & -b_2 t
\end{array}\right),
\label{threeH1}
\ee
where $b_1\,, b_2,\, \epsilon,\, g_{12},\, g_{23},\, g_{13}$ are free parameters, and we allow complex values of the off-diagonal couplings. All other  MLZ models  with $n=3$ are obtained  from (\ref{threeH1}) by time  shifts $t\rar t+t_0$ and gauge transformations $H\rar  H+(bt+c)\hat{1}_3$, which do not change the transition probabilities, and can be trivially included after our analysis of  model (\ref{threeH1}) \cite{be}. 
This model is found frequently in research, for example, on Bose condensates \cite{bose-3lev}, photoionization \cite{photo-3lev}, physics of neutrino \cite{neutrino-LZ}, and Landau-St\"uckelberg interferometry \cite{kiselev-3lev}.

\subsection{Commuting partner}
Comparing (\ref{threeH1}) and (\ref{mlz2}), we find 
\be
A(\tau)=\left( \begin{array}{ccc}
0 & g_{12} & g_{13} \\
g_{12}^* & \epsilon & g_{23} \\
g_{13}^* & g_{23}^* & 0
\end{array}\right), \quad B(\tau)=\left( \begin{array}{ccc}
b_1 & 0& 0 \\
0 & 0 & 0 \\
0 & 0 & -b_2 
\end{array}\right),
\label{ab-3}
\ee
where all parameters can be functions of the second time variable $\tau$.

Substituting (\ref{ab-3}) into (\ref{ccond1}), we find that
\be
\frac{|g_{ij}(\tau)|^2}{B_{ii}(\tau)-B_{jj}(\tau)}=|\gamma_{ij}|^2, \quad i\ne j =1,2,3,
\label{cond1-3}
\ee
where $\gamma_{12},\, \gamma_{13}, \, \gamma_{23}$ are $\tau$-independent constants. This can be achieved if we assume that $b_1$ and $b_2$ in (\ref{ab-3}) are arbitrary functions of $\tau$, whereas the off-diagonal couplings depend on $\tau$ as
\ben
\nonumber g_{12}& =&\gamma_{12}\sqrt{ b_1(\tau)/\beta_{1}}, \quad g_{23}=\gamma_{23}\sqrt{ b_2(\tau)/\beta_{2}}, \\
\label{gg-3}
g_{13}&=&\gamma_{13}\sqrt{(b_1(\tau)+b_2(\tau))/(\beta_{2}+\beta_{1})}, \\
\nonumber \beta_{1} &\equiv& b_1(1), \quad \beta_2 \equiv b_2(1),
\een
where we introduced the parameters $\beta_{1,2}$ to set, for convenience, that  $g_{ij}=\gamma_{ij}$ at $\tau=1$.

Note that  condition (\ref{cond1-3}) has a simple meaning. Namely, for model (\ref{threeH1}) to have a nontrivial commuting partner, the dependence of its parameters on $\tau$ should conserve the LZ adiabaticity parameters for all pairs of diabatic levels.
The same property is also found in the fully solvable MLZ models with time-linear commuting partner operators \cite{parallel-LZ}.

For a general three-state matrix $A(\tau)$ with nondegenerate eigenvalues, we can satisfy  the condition (\ref{ccond3}) by assuming that the matrix $D(\tau)$ has the form 
\be
D(\tau) = \frac{r_2(\tau)}{2} A(\tau)^2,
\label{d3}
\ee
where $r_{2}(\tau)$ is, at this stage, an arbitrary function of $\tau$. We could use any other power of $A(\tau)$ in (\ref{d3}) but for a 3$\times$3 matrix the higher powers would reduce to combinations of $A(\tau)$ and $A(\tau)^2$. We also did not find any advantage in adding a linear in $A(\tau)$ term to $D(\tau)$ in this particular example.

The remaining conditions in (\ref{ccond2}) are equivalent to  three coupled equations on all parameters. Substituting (\ref{d3}) into (\ref{ccond2}),  we find that they all are satisfied if 
\ben
\label{inv1}
&&r_2(\tau)=\frac{b_2\partial_{\tau} b_1 - b_1 \partial_\tau b_2}{b_1b_2(b_1+b_2)},\\
\label{inv2}
&&\partial_{\tau} S(\tau)=0, 
\een
where
\be
S(\tau)=\frac{\epsilon^2}{2} \left(
\frac{1}{b_1} +\frac{1}{b_2}
\right).
\label{sdef}
\ee
From Eq.~(\ref{inv2}), we find explicitly that 
\be
\epsilon(\tau) = \epsilon_0 \sqrt{\frac{b_1(\tau) b_2(\tau)}{b_1(\tau) +b_2(\tau)}},
\label{inv24}
\ee
where $\epsilon_0$ is an arbitrary constant parameter.

Summarizing, there is  generally a nontrivial partner operator for the 3-state Hamiltonian (\ref{threeH1}):
\be
H_3'(t,\tau)=\frac{\partial_{\tau}B(\tau)}{2} t^2 +\partial_{\tau}A(\tau)t +\frac{r_2(\tau)}{2} A(\tau)^2,
\label{threeHp1}
\ee
where matrices $B(\tau)$ and $A(\tau)$ 
are given in (\ref{ab-3}), and where only the $\tau$-dependence of $b_1(\tau)$ and $b_2(\tau)$ is arbitrary, whereas the couplings $g_{ij}(\tau)$ and $\epsilon(\tau)$ are given by equations
(\ref{gg-3}) and (\ref{inv2}), and  $r_2(\tau)$ is given by (\ref{inv1}), with 
 free constant parameters $\gamma_{ij}$ and $\epsilon_0$.
Operators $H_3(t,\tau)$ and $H_3'(t,\tau)$   satisfy the integrability conditions (\ref{cond1})-(\ref{cond2}) with $\tau_1=t$ and $\tau_2=\tau$.

\subsection{Area conservation rule}
\label{area-sec}}
Function~$S=S(\tau)$ in (\ref{sdef}) has a simple geometric interpretation.
Consider the plot of diabatic levels, i.e., the time-dependence of the diagonal elements of the Hamiltonian (\ref{threeH1}). Figure~\ref{three-levels-pic} shows that the diabatic levels enclose a triangle whose area is precisely $S(\tau)$ from Eq.~(\ref{sdef}).

Thus we can formulate the condition (\ref{inv2}) as the property that in order to have the integrability partner operator (\ref{threeHp1}), 
the  $\tau$-dependence of  the Hamiltonian (\ref{threeH1}) should be such that the area enclosed by the diabatic levels is independent of $\tau$. This property is an extension of the ``zero area" rule that is found in all known fully solvable MLZ models  \cite{quest-LZ}.

\subsection{Invariant transition probabilities}

\begin{figure}[t!]
    \centering
    \includegraphics[width=0.45\textwidth]{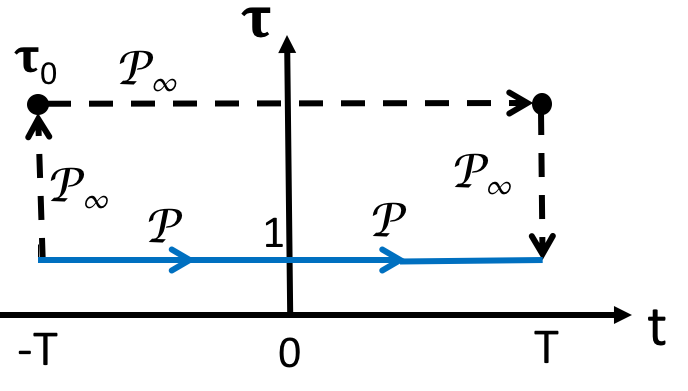}
    \caption{An integration path ${\cal P}$ (blue arrows) with $\tau=1$ and $t\in (-T,T)$ can be deformed into a path ${\cal P}_{\infty}$, such that the horizontal part of ${\cal P}_{\infty}$ has $\tau=\tau_0\ne 1$ (dashed black arrows). Such deformations do not change the evolution matrix. Vertical legs of ${\cal P}_{\infty}$ have $t=\pm T $ with $T\rar \infty$, so they contribute only to the trivial adiabatic phases in the evolution matrix, and do not affect the transition probabilities. For the three-state LZ model, the path ${\cal P}_{\infty}$ can be chosen so that $b_1\rar \infty$ along the horizontal piece of this path. }
    \label{paths-fig}%
\end{figure}

 The precise form of the commuting operator $H_3'$ is not important for our further analysis. 
 What is important for the MLZ problem is that the integrability conditions allow  deformations of the integration path in the two-time space $(t,\tau)$. Namely, let us define the evolution operator
 \be
 U=\hat{\cal T}_{\cal P}\exp \left( -i \int_{\cal P} H_3(t,\tau)\, dt+H_3'(t,\tau) \, d\tau \right),
 \label{path1}
 \ee
 where $\hat{\cal T}_{\cal P}$ is the path ordering operator along  ${\cal P}$ in the two-time space $(t,\tau)$.
  The integrability conditions (\ref{cond1})-(\ref{cond2}) mean that the nonabelian gauge field with components ${\bf A}(t,\tau) = (H_3,H_3')$ has zero curvature. This means in turn that the result of integration in (\ref{path1}) does not change after smooth deformations of ${\cal P}$  that keep only the initial and final points of ${\cal P}$ intact \cite{commute}, and avoid singularities of the $\tau$-dependent Hamiltonians, as we show in Fig.~\ref{paths-fig}.

 Let the physical evolution correspond to changes of $t$ from $-\infty$ to $+\infty$ at $\tau=1$. Then, ${\cal P}$ starts at the point $(t,\tau)=(-\infty,1)$.  We are free to fix, initially, $t$ and change $\tau$ from this point to another value, and only then perform $t$-evolution at fixed new $\tau$. After this, we  can bring $\tau$ back to $\tau=1$ at $t=+\infty$ \cite{commute,faddeev-book}. 
 
 The $\tau$-evolution at fixed $t=- \infty$ or $t=+\infty$ is purely adiabatic due to the quadratic dependence of the diagonal elements of $H_3'$ on $t$. Therefore, the transition probability matrices of the 3-state LZ-models that differ  only by  $\tau$, are identical.
 
\begin{figure}[t!]
    \centering
    \includegraphics[width=0.45\textwidth]{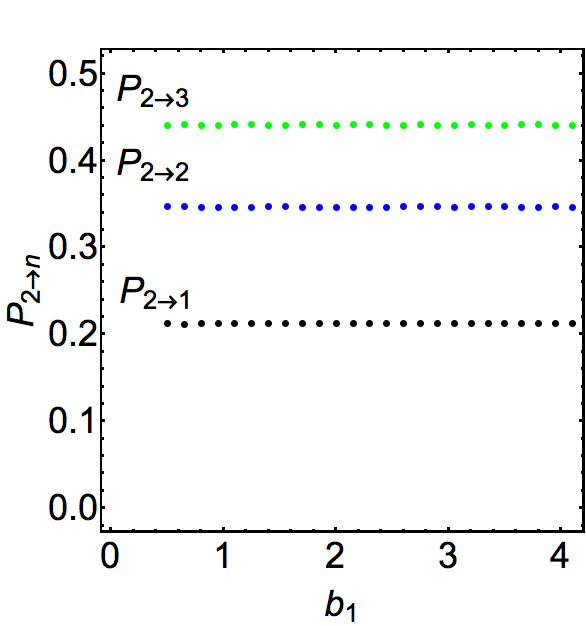}
    \caption{The numerical  check for the  transition probability independence of the level slope $b_1$ in  model (\ref{threeH1}). Here, we assume that $b_1=\tau$ and $b_2={\rm const}$. The other parameters of this model depend on $\tau$ according to Eqs.~(\ref{gg-3})~and~(\ref{inv24}). The fixed parameters for the simulations were $b_2=1$, $\gamma_{12}=0.354$, $\gamma_{23}=0.3$, $\gamma_{13}=0.327$, $\epsilon_0=0.52$. The evolution starts from the 2nd diabatic state (whose diabatic level has zero slope).  }
    \label{equal-three}%
\end{figure}

Thus, if we change $b_1(\tau)$ and $b_2(\tau)$  almost arbitrarily and adjust the other parameters according to (\ref{gg-3}) and (\ref{inv24}), then the transition probabilities in the model (\ref{threeH1}) do not change. The only condition on the possible parameter deformations is that they must  be continuous, such that the diabatic level slopes remain nondegenerate. Figure~\ref{equal-three} shows our results of the  numerical solution of the time-dependent Schr\"odinger equation with the Hamiltonian (\ref{threeH1}). We fixed $b_2$ to a constant and assumed that $b_1=\tau$, i.e., we identified $\tau$ with the slope of the first diabatic level. This figure shows that, indeed, the variations of $b_1$, with adjustments of the other parameters according to (\ref{gg-3}) and (\ref{inv2}), did not change the transition probabilities.

This  is the most general effect of the integrability in the sense of Eqs.~(\ref{cond1})-(\ref{cond2}). Generally, such conditions do not lead to the possibility to write the scattering matrix in terms of known special functions. Rather, the integrability means that there is a possibility to identify a parameter $\tau$ whose changes do not change the transition probabilities and only trivially change the phases of the scattering matrix elements.

\subsection{Reduction of the order of the evolution equation} 
\label{reduction-sec}
The independence of  the transition probabilities of $\tau$  simplifies the analysis of the scattering amplitudes.
In what follows, for simplicity of notation, we will assume that all $g_{ij}$ are purely real. Generalization to complex $g_{ij}$ is straightforward.

\subsubsection{Time-scale separation as $b_1\rar \infty$}

As in the example in Fig.~\ref{equal-three}, let us identify $\tau$ with the slope $b_1$: $b_1=\tau$, with the couplings given by (\ref{gg-3}). We  consider  the limit $b_1\rar \infty$. A simplification follows then because the nonadiabatic transitions between two diabatic states $i$ and $j$ happen during the effective time 
\be
\delta t \sim g_{ij}/(B_{ii}-B_{jj}).
\label{teff}
\ee
Our $\tau$-transformations, given by (\ref{gg-3}) do not change the dimensionless combination, $\gamma_{ij}=g_{ij}^2/(B_{ii}-B_{jj})$. Hence, in the limit $b_1\rar \infty$ the time for nonadiabatic transitions to and from  level 1 is vanishing in comparison to such transitions between levels 2 and 3. 

Thus, as far as the interactions of level 1 are concerned,  for $b_1 \rar \infty$, we can disregard the coupling $g_{23}$ between levels 2 and 3, and disregard  slopes of these levels in comparison to $b_1$. The effective interaction between the three levels is then described by simplified equations:
\ben
\label{a11}
i\dot{a}_1 &=& b_1 t a_1 +\gamma_{12}\sqrt{b_1} a_2 +\gamma_{13} \sqrt{b_1} a_3,\\
\label{a21}
i\dot{a}_2&=&\gamma_{12} \sqrt{b_1} a_1,\\
\label{a31}
i\dot{a}_3 &=&\gamma_{13} \sqrt{b_1} a_1.
\een
It is convenient now to introduce two orthogonal states:
\be
|\psi_1\ra=\frac{\gamma_{12}|2\ra + \gamma_{13}|3\ra}{\sqrt{\gamma_{12}^2+\gamma_{13}^2}}, \quad
|\psi_2\ra=\frac{\gamma_{13}|2\ra - \gamma_{12}|3\ra}{\sqrt{\gamma_{12}^2+\gamma_{13}^2}}.
\label{psis}
\ee
We can then introduce their amplitudes, so that 
\be
\left( \begin{array}{c}
a_2\\
a_3
\end{array} \right)=\alpha_1(t) |\psi_1\ra +\alpha_2(t) |\psi_2 \ra.
\label{vec1-11}
\ee
Substituting (\ref{vec1-11}) into (\ref{a21})-(\ref{a31}), we find that 
\be
\alpha_2={\rm const},
\label{psi2}
\ee
and 
\ben
\label{a31-1}
\nonumber i\dot{a}_1 &=& b_1 t a_1 +\gamma \alpha_1 ,\\
i\dot{\alpha}_1&=&\gamma a_1, 
\een
where 
\be
\gamma\equiv \sqrt{b_1(\gamma_{12}^2+\gamma_{13}^2)}.
\label{def:gamma}
\ee
Equations (\ref{a31-1}) are the evolution equations for amplitudes in the two-state Landau-Zener (LZ) model, whose scattering matrix is known in analytical form.

\subsubsection{Initially empty level 1}

If  level 1 is initially empty, then the survival amplitude $\alpha_1(t=+\infty)$ for  Eqs.~(\ref{a31-1}) is given by the exact LZ formula:
\be
\alpha_1(t)_{t\rar \infty} = e^{-\pi(\gamma_{12}^2 +\gamma_{13}^2)} \alpha_1(-t)_{t\rar \infty}. 
\label{lz1}
\ee
Let us define the evolution operator $U_0$ such that
\be
\left( \begin{array}{c}
a_2(t)\\
a_3(t)
\end{array} \right)_{t\rar \infty} =U_0\left( \begin{array}{c}
a_2(-t)\\
a_3(-t)
\end{array} \right)_{t\rar \infty}.
\label{vec1}
\ee
From (\ref{vec1-11}), (\ref{psi2}), and (\ref{lz1}) we find that 
\be
U_0 = e^{-\pi(\gamma_{12}^2 +\gamma_{13}^2)} |\psi_1 \ra \la \psi_1| + |\psi_2 \ra \la \psi_2|.
\label{u0}
\ee
Knowledge of $U_0$ is not sufficient to find the transition probabilities in the original three-state model because $U_0$ defines the scattering due to fast interactions with level 1. The value $t= +\infty$ in (\ref{vec1}), is actually a time that is much larger than $\delta t \sim \gamma/b_1$ but much smaller than the time scale for other interactions between levels 2 and 3. 

For much longer times than the interval  $\delta t \sim \gamma/b_1$ around $t=0$, where $b_1\rar \infty$ and $\gamma$ is defined in (\ref{def:gamma}), the diabatic states $|2\ra$ and $|3\ra$ interact with each other via two channels. First, the Hamiltonian (\ref{threeH1}), even for  $b_1 \rar \infty$, has a finite coupling in the subspace of the states $|2\ra$ and $|3\ra$:
\be
H_{23}^0 = \left(\begin{array}{cc}
\epsilon_0 & g_{23}\\
g_{23} & -b_2 t
\end{array}\right),
\label{h_eff1}
\ee
where $\epsilon_0$  is given by Eq.~(\ref{inv24}) according to the area conservation, of the triangle in Fig.~\ref{three-levels-pic}, after changes of $b_1$ from its initial value to $b_1\rar \infty$.
In addition, the virtual interactions of levels 2 and 3 with level 1 happen even long after the termination of the nonadiabatic transitions to level 1. The virtual transitions renormalize the couplings and the diabatic energies of  levels 2 and 3. This effect on the effective two-state Hamiltonian is needed  up to the zeroth order in  $1/b_1$, and can be calculated perturbatively.  

\begin{figure}[t!]
    \centering
    \includegraphics[width=0.45\textwidth]{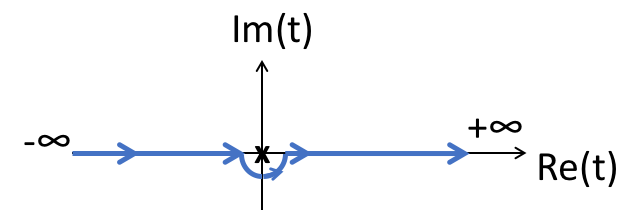}
    \caption{The integration path for the evolution with the Hamiltonian (\ref{heff23}) that provides the transition amplitudes between levels 2 and 3 in the original three-state LZ model (\ref{threeH1}) when the level 1 is initially empty.    }
    \label{paths}%
\end{figure}

To include this renormalization we  recall that the amplitudes of the diabatic states are analytic functions of time. If we integrate over the effect of the 1st level, the fast virtual transitions throughout the 1st level can be included by adding a
$\sim 1/t$ term to the effective Hamiltonian of states 2 and 3:
\be
H_{23}(t)=H_{23}^0(t) -\frac{\gamma^2}{b_1 t}|\psi_1\ra \la \psi_1|,
\label{heff23}
\ee
and assuming that the integration over time happens along the path shown in Fig.~\ref{paths}. Note that $\gamma \sim \sqrt{b_1}$, so the last contribution to (\ref{heff23}) does not depend on $b_1$ explicitly.

The integration path encloses the point $t=0$ from an infinitesimally small distance in the lower complex half plane, as shown in Fig.~\ref{paths}. 
This small deviation of the time into the complex plane produces 
the exponential prefactor in Eq.~(\ref{u0}). The $\sim 1/t$ interactions along the real time are then needed to keep the causality of the response to this perturbation \cite{kk-sinitsyn}. 

The transition probabilities between levels $2$ and $3$ are found then from the evolution matrix with the Hamiltonian (\ref{heff23}) during the time interval $(-\infty,+\infty)$. To find the transition probability from level 2 or 3 to level 1, we should find the scattering probability from the initial level as $t\rar -\infty$ to $|\psi_1\ra $ as $t\rar 0_{-}$, and then multiply this probability by  $1-e^{-2\pi \gamma^2/b_1} $, which is the LZ probability to turn to  level 1 during a short time interval near $t=0$.

\subsubsection{Initially filled level 1}
\label{filled-1-sec}

For the initial conditions 
with $a_1(-\infty) = 1$, $a_2(-\infty) = a_3(-\infty) = 0 $, Eqs.~(\ref{a21}) and (\ref{a31}) lead to 
$$
a_3(t) = \frac{\gamma_{13}}{\gamma_{12}} a_2(t).
$$
Substituting this into (\ref{a11}) and introducing a new variable 
$$
a_2'=\frac{\sqrt{\gamma_{12}^2 + \gamma_{13}^2} a_2}{\gamma_{12}},
$$
we find the  standard two-state LZ model:
\ben
\label{a12}
\nonumber i\dot{a}_1 &=& b_1 t a_1 +\gamma a'_2,\\
\label{a22}
i\dot{a}'_2&=& \gamma a_1,\\
\een
where $\gamma$ is defined in (\ref{def:gamma}).

The LZ probability to remain on the 1st level as $t\rar +\infty$ is then given by
\be
P_{1\rar 1} =e^{-2\pi (\gamma_{12}^2 +\gamma_{13}^2)}.
\label{p11-1}
\ee
This result is exact but well known. Since level 1 has the highest slope among the three diabatic levels of the original three-state system, the survival probability for the corresponding diabatic state is given by 
the Brundobler-Elser formula \cite{be}, 
which applies to any MLZ system and which in our case  produces the probability (\ref{p11-1}).

More interesting consequences are found for the transition probabilities to the other states. After the direct fast transitions from level 1 terminate, i.e. near $t=0_+$, the system  is in the state $|\psi_1 \ra$ with the probability
$$
P_{1\rar \psi_1} = 1-P_{1\rar 1}=1-e^{-2\pi \gamma^2/b_1}.
$$
Then, $|\psi_1 \ra$ becomes the initial state for the ``slow" evolution in the subspace of states $|2\ra$ and $|3\ra$ with the Hamiltonian $H_{23}$ from Eq.~(\ref{heff23}). State $|1\ra$  is then considered completely decoupled.  Let $P_{\psi_1 \rar 2}$ be the transition probability for the following evolution 
from $t\rar 0_+$ to $t\rar +\infty$. Then, the transition probability $P_{1\rar 2}$ in the original three-state model is given by
\be
P_{1\rar 2}=P_{1\rar \psi_1} P_{\psi_1 \rar 2},
\label{P12-fin}
\ee
and there is an analogous formula for $P_{1\rar 3}$:
\be
P_{1\rar 3}=P_{1\rar \psi_1} P_{\psi_1 \rar 3}.
\label{P13-fin}
\ee

Summarizing this section, we found that  the transition probabilities in the original three-state model can be found from the scattering probabilities in a simpler two-state problem with the Hamiltonian (\ref{heff23}). If the latter problem is solvable then the original MLZ problem is solvable too. This happens for $\epsilon_0=0$ and $g_{13}=0$. Then, the Schr\"odinger equation with the Hamiltonian 
(\ref{heff23}) reduces to the confluent hypergeometric equation, whose solution has known asymptotic behavior. Indeed, the original three-state LZ model for $\epsilon_0=0$ has a known exact expression for the transition probabilities.

Unfortunately, if $\epsilon_0 \ne 0$, the two-state model (\ref{heff23}) does not reduce to a hypergeometric equation. For a 2nd order ordinary differential equation, whose coefficients have  simple time-dependence, this means that the problem is likely not  solvable  analytically.
In turn, this means that there is likely no  analytical solution of the general three-state LZ model either. However, the reduction of the complexity from a 3rd order to  a 2nd order differential equation leads to considerable simplifications for asymptotic analysis.

\section{The  transition probabilities in the adiabatic limit}
\label{adiab-sec}

Two-state scattering problems for evolution between $t=\pm \infty$, with an explicitly time-dependent Hamiltonian, have a general semiclassical  solution. In the adiabatic limit, this solution becomes asymptotically exact. Namely, the Dykhne formula \cite{approx-LZ5} provides the over-gap transition probability:
\be
P_D=e^{-2 {\rm Im} \left[ \int_{t_0}^{t_1} (E_+ (t) - E_{-}(t)) \, dt \right]},
\label{dykhne1}
\ee
where $E_{\pm}$ are the larger/smaller eigenvalues of the 2$\times$2 Hamiltonian, $t_0$ is any time point on the real time axis, and
$t_1$ is the value of time in the upper half of the complex plane for which 
$$
E_+ (t_1)=E_{-} (t_1).
$$
The integration path in (\ref{dykhne1}) goes so that the imaginary part of the  integral is always growing along this path.
Generally, there are many  such time points. In order to obtain the leading exponent, one has to find $t_1$ for which the integral in (\ref{dykhne1}) has the smallest value. The Dykhne formula's prediction for the over-gap transition probability in the 
two-state LZ model coincides with the prediction of the exact LZ model's solution. Generally, for the nearly-adiabatic evolution, the Dykhne formula produces correct predictions only for the dominating exponentially small contribution to the over-gap transition probability, including the leading unit prefactor at this exponent. 

Interestingly, for the three-state scattering problem, there is no general analog of the Dykhne formula. This difficulty  was noticed in the analysis of \cite{approx-LZ0}. There have been  efforts to solve this problem \cite{approx-LZ1,approx-LZ2,approx-LZ3,approx-LZ4} but with only  a limited progress. Nevertheless, our finding that the three-state LZ problem is reducible to a 2nd order scattering problem can be used to estimate the transition probabilities using  (\ref{dykhne1}). For illustration, we consider a simple case: 
\be
H_3=\left( \begin{array}{ccc}
b t & g & 0\\
g & \varepsilon/\sqrt{2} & g \\
0 & g & -bt
\end{array}
\right),
\label{threeH-demo1}
\ee
where we assume that the parameters satisfy the adiabaticity condition:
\be
g^2/b \gg 1,
\label{adiab-c}
\ee
and we restrict this example only to  
$$
\varepsilon>0, \quad b>0.
$$

\begin{figure}[t!]
    \centering
    \includegraphics[width=0.45\textwidth]{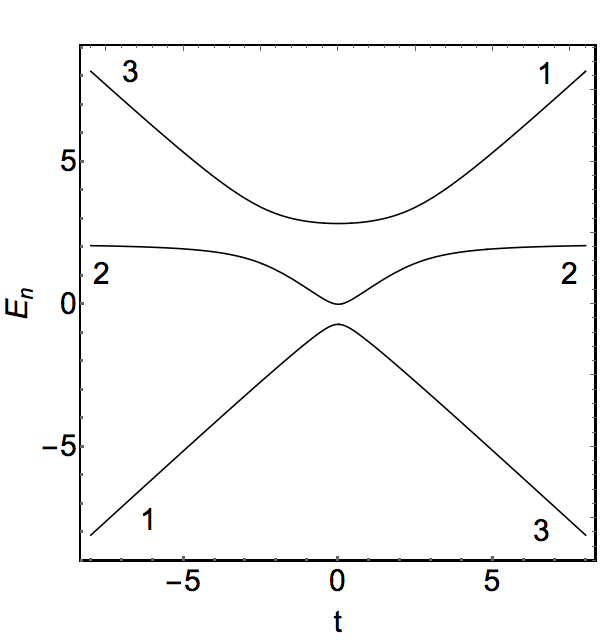}
    \caption{The time-dependent eigenenergies of the Hamiltonian  (\ref{threeH-demo1}). The numbers mark the diabatic levels, with which the eigenenergies merge as $t \rar \pm \infty$. The gaps between diabatic levels 1 and 2, as well as 2 and 3, are induced by direct interactions between the corresponding states. The gap between levels 1 and 3 is smaller because it is induced by indirect interactions.
    The choice of parameters: $b=1$, $g=1$, $\varepsilon=3$. }
    \label{three-spectrum}%
\end{figure}
In Fig.~\ref{three-spectrum}, we show the time-dependent spectrum of the Hamiltonian (\ref{threeH-demo1}). The diabatic levels 1 and 3 do not couple
to each other directly but a gap 
between the corresponding eigenenergies appears due to higher order interactions. Transitions through such gaps, which emerge due to state couplings via virtual transitions through other levels, are particularly hard to take into account during semiclassical calculations. Such situations, however, become often physically important \cite{bose-3lev,anglin}.

 For  $\varepsilon<g $, all gaps are of the order of $g$. Condition (\ref{adiab-c}) leads then to the adiabatic evolution, such that the system transfers from level 1 to level 3 with almost unit probability. However, as $\varepsilon$ increases, the gap between diabatic levels 1 and 3  diminishes. Hence, for $\varepsilon \gg g$, the system passes on level 1 throughout the gap with level 3 nonadiabatically and then remains on level 2. Thus, 
as $\varepsilon$ increases, the transition probability from level 1 to level 2 changes from almost $0$ to almost $1$.

Let us quantify this process. We are interested in the transition probability $P^3_{1\rar 2}$, in which the upper index $3$ means that this probability describes the evolution with the three-state Hamiltonian (\ref{threeH-demo1}).
Following previous subsection, we find for our case that $|\psi_1\ra =|2\ra$. The problem reduces then to finding the over-gap scattering probability for the evolution with the Hamiltonian
\be
H_2(t)=\left( \begin{array}{cc}
 -\kappa/t + \varepsilon &g \\
g & -bt
\end{array}
\right), \quad \kappa = g^2/b.
\label{twoH}
\ee
Note the disappearance of $\sqrt{2}$ in the denominator at $\varepsilon$ in comparison to (\ref{threeH-demo1}), which follows from the  area conservation rule (section~\ref{area-sec}).
Let us mark the diabatic states of $H_2$ by indices $2$ and $3$ in order to keep their relations to the corresponding states of $H_3$. We also define the probability $P_{2\rar 2}$  to be in the state $|2\ra$ of this two-state model as $t\rar +\infty$ given that as $t\rar 0_+$, the system starts  in this state too. Since the system passes through an avoided level crossing, $P_{2\rar 2}$ is also the probability of the over-gap transition for the evolution with the Hamiltonian (\ref{twoH}), so, we can apply the Dykhne formula to describe it.

The evolution over $t\in(0_+,+\infty)$ can be transformed into the evolution over  $s \in (-\infty,\infty)$ after a change of variables
$$
 t\sim e^{s},
$$
which sets the problem to the standard form for application of the Dykhne formula. However, for $t_0>0$ in (\ref{dykhne1}), this change of variables is not needed because it leads to the same final expression as to what we get if we apply the Dykhne formula to the Hamiltonian $H_2(t)$ right away. Hence, we will set  $t_0=1$, apply the Dykhne formula without the change of variables, and verify the validity of our results at the end numerically.

The difference of the eigenvalues of $H_2(t)$ is
\be
\Delta E \equiv E_+-E_{-} = \sqrt{(\varepsilon  -\kappa/t + bt)^2+4g^2}.
\label{en-diff1}
\ee

The points with branch cuts are given by 
\be
\varepsilon -\kappa/t + bt = \pm i 2g, \quad \kappa=g^2/b.
\label{roots}
\ee
This is a quadratic equation on $t$, so its roots in the upper complex plane can be found analytically:
\be
t_{1,2}=\frac{2ig -\varepsilon \pm \sqrt{\varepsilon^2-4i\varepsilon g}}{2b}.
\label{roots1}
\ee
\begin{figure}[t!]
    \centering
    \includegraphics[width=0.45\textwidth]{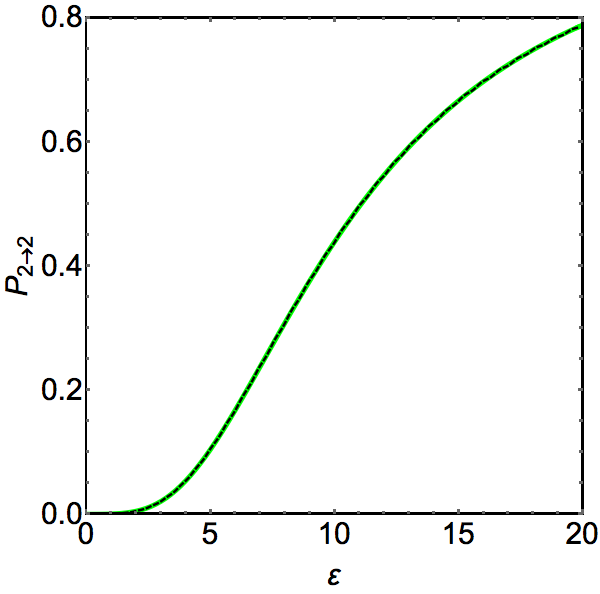}
    \caption{The Dykhne formula's prediction (green solid curve) for the over-gap transition probability in the model (\ref{twoH}) at $g=2$, $b=1$, $\kappa=g^2/b$, $\eta=1$. The black dashed curve is the result of numerically exact solution of the time-dependent Schr\"odinger equation during the time interval $t\in (-1000,1000)$.}
    \label{ncheck1}
\end{figure}

We verified numerically (Fig.~\ref{ncheck1}) that if we take the route $t_1$, and then substitute it into (\ref{dykhne1}) with a trivial integration path that connects $t_1$ and $t_0=1$ along a straight line, the Dykhne formula  describes the desired transition probability correctly. The deviations do appear (they are not visible in Fig.~\ref{ncheck1}) but only for small $\varepsilon$ and only for the 
exponential  prefactor, which remains of the order O(1). That is, the desired transition probability is
\be
P_{2 \rar 2} =\eta e^{-2 {\rm Im} \left[\int_1^{t_1} \Delta E  \, dt \right]},
\label{Pfin1}
\ee
where $\eta$ is the prefactor.


We attribute the difference of numerically found  $\eta$ from unity, for small $\varepsilon$,  to two facts. First, the semiclassical theory applies to the domain $g^2/b \gg 1$. In our simulations, $g^2/b = 4$ was not made very large in order to keep the desired calculation precision. Second, at $\varepsilon=0$ the routes $t_1$ and $t_2$ 
coincide. Hence, this is where our  approach has to be upgraded. It is not surprising then that for finite but small $\varepsilon$ subdominant exponents produce visible contributions.

In fact, the case with $\varepsilon=0$ is exactly solvable \cite{sinitsyn-14pra}, and predicts that
\be
P_{2\rar2}(\varepsilon=0) = \frac{2e^{-\pi g^2/b}}{1+e^{-\pi g^2/b}}.
\label{p22-exact}
\ee
The standard perturbative approach in small $\varepsilon$ near this exact solution can be developed but this would be beyond our goals here. We  note also that such a higher order degeneracy of the branch cut is described by the semiclassical theory of A. Joye \cite{joye-prefactor} that predicts $\eta=2$  prefactor at the leading exponent for $\varepsilon=0$, which is confirmed by the exact solution. Hence, the semiclassical analysis can be developed in the vicinity of $\varepsilon =0$ too.

\begin{figure}[t!]
    \centering
    \includegraphics[width=0.45\textwidth]{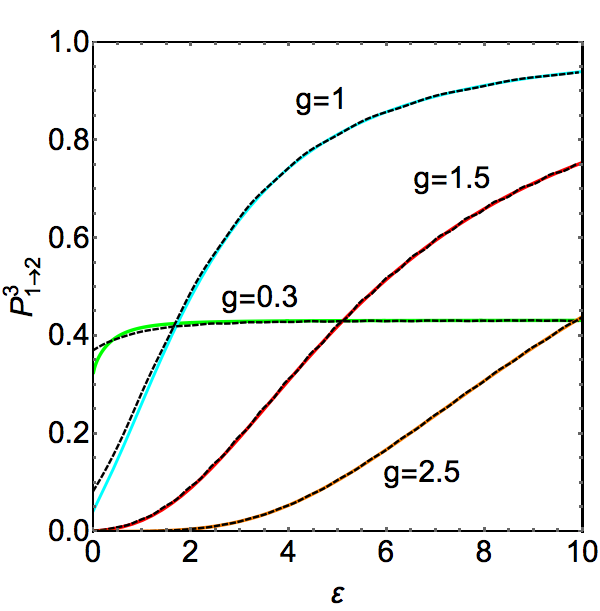}
    \caption{The semiclassical prediction of equation~(\ref{desiredP})  (colored solid curves) for the  probabilities of transitions from the state with the highest energy level slope to the state with zero energy level slope in the model (\ref{threeH-demo1}) with $b=1$,  $\eta=1$,
    and $g=0.3$ (green), $g=1$ (cyan), $g=1.5$ (red) and $g=2.5$ (orange). The black dashed curves are the results of numerically exact solution of the time-dependent Schr\"odinger equation within the time interval $t\in (-1000,1000)$.}
    \label{ncheck3}
\end{figure}
Finally, using the results of section~\ref{filled-1-sec}, we find the desired leading exponent for the transition probability in the three-state LZ model (\ref{threeH-demo1}):
\ben
\label{desiredP}
 && P^3_{1\rar 2} = \left(1-e^{-\frac{2\pi g^2}{b}  } \right) P_{2\rar 2} = \\
\nonumber&& = \left(1-e^{-\frac{2\pi g^2}{b}  } \right) e^{-2 {\rm Im} \left[\int_1^{t_1} \Delta E  \, dt \right]},
\een
where $\Delta E$ and $t_1$ are given by (\ref{en-diff1}) and (\ref{roots1}), respectively.
The inclusion of  $e^{-2\pi g^2/b}$ in (\ref{desiredP}) is, formally, beyond the precision of semiclassical calculations. However, this term is provided by the exact part of our theory, and we found it useful because it produces accurate prediction for $P^3_{1\rar 2}$ even when $g^2/b$ is small, 
as we show in figure~\ref{ncheck3}. 
It is, actually, surprising how accurately the formula (\ref{desiredP}) applies in the whole domain of the parameters with positive $\varepsilon$.

 We provided this example to illustrate that the integrability  leads to the very possibility of such a semiclassical analysis, which   for more than two interacting states  remains generally  undeveloped. In our example, the integrability conditions (\ref{cond1})-(\ref{cond2}) play a role that is similar to how the integrability in classical mechanics helps to study quantum mechanical systems. 
Indeed, if a classical model with a time-independent Hamiltonian is integrable, then there is a straightforward path to explore the quantum version of this model semiclassically using the well-developed WKB approach. 
Analogously, our example suggests that there can  be classes of mutlistate time-dependent problems whose semiclassical description is possible using the integrability conditions (\ref{cond1})-(\ref{cond2}).


We would like also to attract attention to the fact that for $g^2/b>1$ the 
transition probability curve $P_{1\rar 2}^3(\varepsilon)$ in Fig.~\ref{ncheck3} has a sigmoidal shape. Such a dependence of a response function on a parameter is often needed in  memory devices. The transition probability $P_{1\rar 2}^3$ can describe, e.g., an electronic transport  between molecular levels in response to an optical control pulse, whereas $\varepsilon$ can be a parameter that is controlled by  local gate voltage. Thus, the results in Fig.~\ref{ncheck3} suggest that the three-state LZ model (\ref{threeH1}) may describe the work of a quantum memory device structure.

\section{Bipartite models}
\label{bipartite-sec}

Equation~(\ref{ccond2}) is equivalent to $N(N-1)/2 \sim N^2$ coupled constraints on the parameters, where $N$ is the dimension of the Hilbert space. However, the matrix $D(\tau)$ that satisfies (\ref{ccond3}) can generally be written as 
\be
D(\tau) = \sum_{k=1}^{N-1} \frac{r_k(\tau)}{k!} A^k,
\label{dn1}
\ee
with some functions $r_k(\tau)$. 
The number of independent $r_k(\tau)$
is growing as $\sim N$ with the size of the model, as well as the number of diagonal elements of the MLZ Hamiltonian, such as $B_{ii}(\tau)$ and $A_{ii}(\tau)$. This means that not all MLZ models have such a quadratic commuting operator with conditions (\ref{cond1})-(\ref{cond2}). 

Nevertheless, large classes of models do have time-quadratic commuting operators. Let us explore here such a case that describes bipartite interactions. The diabatic states of the bipartite MLZ model split into two disjoint sets, $X_1$ and $X_2$. Only states from the different sets couple to each other directly, that is, for any $a,b \in X_1$ and $c,d \in X_2$, we have $A_{ab}=A_{cd}=0$ but generally $A_{ac} \ne 0$ and $A_{bd}\ne 0$.

\subsection{Integrability of bipartite MLZ models}

Let us introduce the parameters that characterize slope differences between diabatic levels: 
\be
\Delta_{ij} \equiv B_{ii}-B_{jj},  \quad \forall i,j.
\label{delta-def}
\ee

As in the three-state model, condition (\ref{ccond1}) means that 
for $\tau$-dependent $\Delta_{ij}(\tau)$ the couplings must satisfy the conditions 
\be
\frac{|A_{ij}(\tau)|^2}{\Delta_{ij}(\tau)} = |\gamma_{ij}|^2, 
\label{ccond111}
\ee
where $\gamma_{ij}$ are $\tau$-independent.

Looking for $D(\tau)$ in the form 
\be
D(\tau) = -\frac{r}{2} A^2(\tau),
\label{dtau2}
\ee
where $r$ is a constant,
and assuming that $a,b \in X_1$, $c \in X_2$, we find that for bipartite models Eq.~(\ref{ccond2}) leads to the conditions
\be
\label{bi1}
\frac{\partial_{\tau} \Delta_{ac}}{\Delta_{ac}} -\frac{\partial_{\tau} \Delta_{bc}}{\Delta_{bc}}=r(\Delta_{ac} -\Delta_{bc}).
\ee

\subsection{LZ-chains}
Consider a  special case of the MLZ model, which is called the LZ-chain \cite{sinitsyn-chain}: 
\be
\quad A_{ij}=0\quad  \quad \forall \, |i-j|>1.
\label{chain-def}
\ee
It is convenient to denote the nonzero parameters as 
\be
\beta_n\equiv B_{nn},\quad \varepsilon_n\equiv A_{nn}, \quad g_n\equiv A_{n,n+1}.
\ee

For simplicity, here we consider only the case with
\be
\beta_n=\beta n, \quad \varepsilon_n=0, \quad \forall \,n.
\label{beta-def}
\ee
Odd and even $n$-indices correspond to the states  of the two different sets of the bipartite graph.

Let us now introduce functions $\beta_n(\tau)$ such as $\beta_n(0)$ coincide with the slopes $\beta_n$ in (\ref{beta-def}). 
We also denote
\be
\Delta_n = \beta_{n+1}(\tau)-\beta_n(\tau).
\label{deltan-def}
\ee
All $\Delta_n$ are functions of $\tau$ with initial conditions $\Delta_n(0)=\beta$. Conditions  (\ref{bi1}) become
\be
\frac{\partial_{\tau} \Delta_n}{\Delta_n}-\frac{\partial_{\tau} \Delta_{n+1}}{\Delta_{n+1}}=-r(\Delta_n + \Delta_{n+1}).
\label{chain1}
\ee
Equations~(\ref{chain1}) can be solved sequentially. 
Let us denote
$$
q(\tau) \equiv e^{\beta r \tau}.
$$
We find then $\Delta_0=\beta$, and
\begin{eqnarray}
\nonumber \Delta_1&=&\frac{\beta q(\tau)}{2-q(\tau)},\\
\nonumber \Delta_2&=&\frac{\Delta_1}{3-2q(\tau)},\\
\nonumber \Delta_3&=& \frac{2\Delta_2}{4-3q(\tau)} - \Delta_2,
\end{eqnarray}
and so on.
Substituting this into (\ref{deltan-def}) we find a simple pattern that leads to the general expression for the level slopes: 
\be
\beta_n=\frac{\beta n}{n-(n-1)q(\tau)}=\frac{\beta n}{n[1-q(\tau)]+q(\tau)}.
\label{slopes}
\ee
From (\ref{slopes}), we then recover the general solution for $\Delta_n$:
\be
\Delta_n=\frac{\beta q}{[n(q-1)-1][n(q-1)-q]}.
\label{deltan-2}
\ee
The coupling constants then depend on $q=q(\tau)$ as 
\be
g_n(q) =g_n^{q=1} \sqrt{\frac{q}{[n(q-1)-1][n(q-1)-q]}}.
\label{g-deform1}
\ee
Note that the original model is recovered at $q(\tau)=1$. Hence, the models that are obtained by setting $q\ne 1$ can be called q-deformations of the original model.

\subsection{Deformed solvable chain models}

 If  $q(\tau)$ changes continuously, the transformations (\ref{slopes}) and  (\ref{g-deform1}) do not change the transition probabilities. Let us illustrate how this fact can be exploited. Consider a simple model of a molecular bosonic field dissociation into  atomic Bose condensates:
\be
AB\rar A+B,
\label{dissociate}
\ee
where $A$ and $B$ are  different bosonic modes, and $AB$-mode is treated in the mean field approximation as a c-number \cite{sinitsyn-chain}.
The Hamiltonian of this model is 
\be
H=\beta t \hat{a}^{\dagger} \hat{a} +\frac{g}{\sqrt{2}}\left[\hat{a}^{\dagger} \hat{b}^{\dagger}+\hat{a}\hat{b} \right],
\label{dissociate2}
\ee
where $\hat{a}$, $\hat{b}$ are the bosonic annihilation operators. 

The model (\ref{dissociate2}) is fully solvable, e.g., its transition probabilities can be written in terms of simple functions of the model's parameters \cite{sinitsyn-chain}. For example, let us look at the evolution of the bosonic operators in the Heisenberg picture:
\be
i\frac{d\hat{a}}{dt}=\beta t \hat{a} + \frac{g}{\sqrt{2}}\hat{b}^{\dagger}, \quad  i\frac{d\hat{b}^{\dagger}}{dt}= -\frac{g}{\sqrt{2}}\hat{a}.
\label{heizH}
\ee
This equation pair is equivalent to the parabolic cylinder equation, whose solution has a well known asymptotic behavior as $t\rar \pm \infty$  \cite{yurovsky}. 
 In our case, the Hamiltonian conserves the difference of the numbers of $b$ and $a$ bosons, so there are independent invariant sectors of this model. Consider the sector whose vacuum state is 
\be
|0\ra \equiv |0,1\ra,
\label{vac1}
\ee
 which contains no $a$-bosons and one $b$-boson. The diabatic states in this sector are marked by the number of $a$-bosons:
\be
|n\ra \equiv |n,n+1\ra, \quad n=0,1,\ldots.
\label{n-def}
\ee
A particularly simple expression  is found then for the average number, $\la n \ra$, of $a$-bosons as $t\rar +\infty$ if the system starts in the vacuum state $|0\ra$ as $t\rar -\infty$ \cite{yurovsky}.
Namely, the solution of 
(\ref{heizH}) is 
$$
\hat{a}(+\infty) =e^{\pi g^2/(2\beta)} \hat{a}+c \hat{b}^{\dagger},
$$
where $\hat{a}$ and $\hat{b}$ are the initial bosonic operators, and $c$ is a coefficient, such that
$$
|c|^2=e^{\pi g^2/\beta}-1.
$$
The phase of $c$ will not be important. We find then 
\ben
\nonumber  \la n \ra&\equiv& \la 0| \hat{a}^{\dagger}(+\infty) \hat{a}(+\infty) |0 \ra =|c|^2\la 0| \hat{b}\hat{b}^{\dagger} |0\ra=\\
\label{n-av1}
&=&2\left(e^{\pi g^2/\beta}-1 \right).
\een

To generate a new solvable model we note that in the basis (\ref{n-def}) the Hamiltonian has the form of the LZ-chain  
with
$\beta_n=\beta n$, and 
\be
g_n=\frac{g}{\sqrt{2}}\sqrt{(n+1)(n+2)}, \quad n=0,1,2,\ldots.
\label{ggn}
\ee

By applying, to our bosonic Hamiltonian, the transformation (\ref{slopes})-(\ref{g-deform1}) at $q=1/2$,  we find that the deformed Hamiltonian has the form:
\be
H_{[q=1/2]} = \sum_{n\ge0} \left[ \frac{2\beta n t}{n+1}|n\ra \la n| +g \left( |n\ra\la n+1|+ |n+1\ra\la n| \right) \right].
\label{Hq12}
\ee
This  Hamiltonian describes a semi-infinite chain with identical off-diagonal couplings and energy levels whose slopes depend on $n$ nonlinearly. In fact, the slopes saturate as $n\rar \infty$ at 
$\beta_{\infty}=2\beta$. Such features are found in the diagrams of diabatic levels in physically relevant but unsolvable MLZ models \cite{anglin}. Hence, the exact solution of the model (\ref{Hq12}) may be useful for obtaining a physical intuition about the nonadiabatic transitions through  such dense regions in the energy spectrum.
\begin{figure}[t!]
    \centering
    \includegraphics[width=0.4\textwidth]{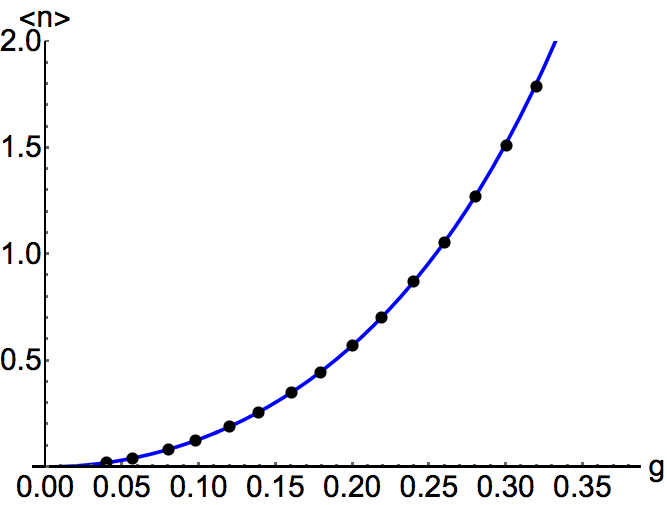}
    \caption{The numerical check of the analytical prediction (the solid blue curve):
    $\la n \ra \equiv \la 0| \hat{n} |0\ra=2\left(e^{\pi g^2/\beta}-1\right)$ as $t\rar +\infty$, where $\hat{n}$ is defined in (\ref{nav-2}). The black points are the results of numerical simulations of the evolution with the Hamiltonian (\ref{Hq12}), which was truncated to the first 12 diabatic states. The simulation time is $t\in(-700,700)$, and
    $\beta=1/2$. The initial state is $|0\ra$ for all parameter choices. }
    \label{n-check-fig}%
\end{figure}

As the $q$-deformations do not change the transition probabilities, the average of the operator 
\be
\hat{n}= \sum_{n=0}^{\infty} n|n\ra\la n|
\label{nav-2}
\ee
at the end of the evolution that starts with the state $|0\ra$ is still given by (\ref{n-av1}). 
In Fig.~\ref{n-check-fig} we show the results of our numerical check of this formula in the context of  model (\ref{Hq12}).
We reemphasize that model (\ref{Hq12}) is a new analytically solvable MLZ model. It became possible to identify  due to the presence of a known $t$-quadratic commuting operator for the LZ-chains.

\subsection{4-state bipartite LZ model}
As another example, we consider a bipartite 4-state model in which levels 1 and 4 interact with levels 2 and 3 with arbitrary couplings between the levels of different groups. 
Equations~(\ref{bi1}) are then given by
\ben
\label{eq4-1}
\nonumber \partial_{\tau} \log \frac{\Delta_{12}}{\Delta_{24}}=r_2(\Delta_{12} +\Delta_{24}),\\
\label{eq4-2}
\nonumber \partial_{\tau} \log \frac{\Delta_{13}}{\Delta_{34}}=r_2(\Delta_{13} +\Delta_{34}),\\
\label{eq4-3}
\nonumber \partial_{\tau} \log \frac{\Delta_{12}}{\Delta_{13}}=r_2(\Delta_{13} -\Delta_{12}),\\
\label{eq4-4}
 \partial_{\tau} \log \frac{\Delta_{24}}{\Delta_{34}}=r_2(\Delta_{24} -\Delta_{34}).
\een
Using that $\Delta_{12} +\Delta_{24}=\Delta_{13} +\Delta_{34}$ and $\Delta_{13} -\Delta_{12}=\Delta_{24} -\Delta_{34}$, we find that 
\be
\Delta_{12}\Delta_{34}=c\Delta_{13}\Delta_{24},
\label{constr4-1}
\ee
where $c$ is a constant. Also, from the definition of $\Delta_{ij}$, we find
\be
\Delta_{13}-\Delta_{12}=\Delta_{24}-\Delta_{34}.
\label{constr4-2}
\ee
If we set one of the slope differences to be a constant, e.g.,
\be
\Delta_{34}= b ={\rm const},
\label{constr4-3}
\ee
then we only should solve the last  equation from (\ref{eq4-4}), which is straightforward. Thus, we find
\begin{widetext}
\ben
\label{dd4-1}
\Delta_{34}&=&b,\\
\label{dd4-2}
\Delta_{24}&=&b(b+b_1)/[b+b_1(1-e^{r_2b\tau})],\\
\label{def:d13}
\Delta_{13}&=&-b(b+b_1)e^{r_2b\tau}/[b(e^{r_2b\tau}-2) +b_1(e^{r_2b\tau}-1)],\\
\label{def:d12}
\Delta_{12}&=&-b^3e^{r_2b\tau}/[(b+b_1(1-e^{r_2b\tau}))(b(e^{r_2b\tau}-2) +b_1(e^{r_2b\tau}-1))],
\een
\end{widetext}
where $b_1$ is another constant, such that $\Delta_{24}(\tau=0)=b_1+b$.

\subsection{Solvable 4-state model with degeneracy}

Let us now introduce a fully solvable four-state model. Imagine that levels $2$ and $3$ from the previous example are degenerate:
\be
H_4=\left(
\begin{array}{cccc}
bt & g_1 & g_2 & 0 \\
g_1^* & 0 & 0 & g_3^* \\
g_2^* & 0 & 0 & g_4^* \\
0 & g_3 & g_4 & -b t
\end{array}
\right),
\label{h4-sol1}
\ee
where we allow all couplings to be complex-valued. 
For such a degeneracy, the diabatic states $2$ and $3$ do not coincide with the adiabatic states as $t\rar \pm \infty$. However, the scattering probabilities between levels $1$ and $4$ are well-defined, as well as the scattering probabilities from $1$ or $4$ to  subspace of states $2$ and $3$. The model (\ref{h4-sol1}) may appear physically when two spins $s=1/2$  interact equally with a linearly time dependent field acting along $z$-axis. Then, the diabatic states $|\uparrow \downarrow \ra$ and $| \downarrow \uparrow \ra$
 are  represented  in (\ref{h4-sol1}) by levels $2$ and $3$ and their diabatic energies are degenerate.
The states $|\uparrow \uparrow \ra$ and $|\downarrow \downarrow \ra$ experience the Zeeman splitting, which corresponds to the linear time-dependence of the diabatic energies of levels $1$ and $4$. The off-diagonal couplings in (\ref{h4-sol1}) then represent various interactions that flip one spin.  Thus, we assume no direct coupling between $|\uparrow \uparrow \ra$ and $|\downarrow \downarrow \ra$. Different coupling values can be the results of spin orbit interactions and   a constant transverse magnetic field, which may act differently on the different spins. Similar MLZ model has previously appeared in studies of molecular nanomagnets (see Fig.~1 in \cite{nanomag}), although the existence of its analytical solution during that time was not known.

The $\tau$-deformations  keep the transition probabilities intact, and our choice of the Hamiltonian (\ref{h4-sol1}) corresponds to $b_1=0$ in (\ref{dd4-2})-(\ref{def:d12}). We set $e^{r_2b\tau} \rar 2$. According to (\ref{def:d13}) and (\ref{def:d12}), this makes  level $1$ have an infinite slope. We then choose a new basis in the degeneracy subspace:
\be
\label{psi}
|\psi_1 \ra = \frac{g_1 |2\ra +g_2 |3\ra}{\sqrt{|g_1|^2+|g_2|^2}}, \quad |\psi_2 \ra = \frac{g_2 |2\ra -g_1 |3\ra}{\sqrt{|g_1|^2+|g_2|^2}}.
\ee

Following the steps in section~\ref{reduction-sec} for the reduction of the order of the Schr\"odinger equation, we integrate out the level with the infinite slope. We find then that in the basis of states $|\psi_1 \ra, |\psi_2\ra, |4 \ra $, the effective Hamiltonian is the time-dependent 3$\times$3 matrix: 

\be
H_3=\left(
\begin{array}{ccc}
-\kappa/t& 0 & \gamma_1 \\
0 & 0 &  \gamma_2 \\ 
\gamma_1^* &  \gamma_2^* & -bt
\end{array}
\right),
\label{ham-3lev1}
\ee
where 
\begin{eqnarray}
\label{kap-def}
 \kappa &\equiv& (|g_1|^2+|g_2|^2)/b, \\
\gamma_1 &\equiv& \frac{g_1g_3^* +g_2g_4^*}{\sqrt{|g_1|^2+|g_2|^2}}, \quad 
\label{gam12-def}
 \gamma_2 \equiv \frac{g_2g_3^* -g_1g_4^*}{\sqrt{|g_1|^2+|g_2|^2}} .
\end{eqnarray}

The probability of a transition $P_{1\rar 4}$ is given by the LZ probability, $1-e^{-2\pi \kappa}$, to transfer from level $1$ to  $|\psi_1\ra$, times the probability $P_{\psi_1 \rar 4}$ of a transition from  $|\psi_1\ra$ as $t\rar 0_+$ to level $4$ as $t\rar +\infty$ during the evolution with the Hamiltonian $H_3$. 

In order to find the latter probability we should solve the model (\ref{ham-3lev1}). Fortunately, this model is exactly solvable. 
It is a special case of the model that was explored in \cite{lin-coulomb} in detail. The model in \cite{lin-coulomb} describes the interaction of a single level with a Coulomb band, in which diabatic energies change with time according to $\sim k_i/t$ law. In our case, $k_1=-\kappa$, and $k_2=0$.
The transition probabilities  can be read directly from the solution of the three-state model in \cite{lin-coulomb}, namely, let
\be
\kappa_{\pm}=(|\gamma_1|^2 \pm |\gamma_2|^2)/b,
\label{kappapm-def}
\ee
and 
\be
s_{\pm} =\frac{\kappa_{+}+\kappa \pm \sqrt{\kappa_+^2 +\kappa(\kappa-2\kappa_{-})}}{2}. 
\label{spm-def}
\ee
The exact solution of the model (\ref{ham-3lev1}) predicts \cite{lin-coulomb}: 
\be
P_{\psi_1 \rar 4} =    \frac{e^{-\pi (\kappa_+ +\kappa)} (e^{\pi s_+}-1)(e^{\pi s_{-}}-1)}{1-e^{-2\pi k}}.
\label{p01}
\ee
Returning to the model (\ref{h4-sol1}), the probability of the transition from level 1 to level 4 is 
\be
P_{1 \rar 4} = \left(1-e^{-2\pi k} \right) P_{\psi_1 \rar 4} =  e^{-\pi (\kappa_+ + \kappa)} (e^{\pi s_+}-1)(e^{\pi s_{-}}-1),
\label{p14-fin}
\ee
where $\kappa$, $\kappa_{\pm}$, and $s_{\pm}$ are defined in (\ref{kap-def}), (\ref{kappapm-def}), and (\ref{spm-def}), respectively.
 
\begin{figure}[t!]
    \centering
    \includegraphics[width=0.45\textwidth]{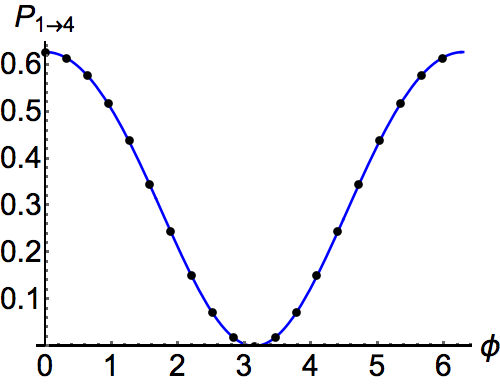}
    \caption{The numerical check of the exact solution (\ref{p14-fin}). The blue solid curve is the analytical prediction for the transition probability $P_{1\rar 4}$ versus the phase $\phi \in (0,2\pi)$ of one of the coupling constants. The black dots are the results of the numerical simulations of  evolution with the 4$\times$4 Hamiltonian (\ref{h4-sol1}).  The choice of the parameters is $b=1$, $g_1=ge^{i\phi}$, $g_2=g_3=g_4=g$, where $g=0.5$.  }
    \label{p14-check-fig}%
\end{figure}

In addition to the probability (\ref{p14-fin}), the Brundobler-Elser formula provides the probability to remain on the energy level with the largest slope \cite{be}:
$$
P_{1\rar 1} = e^{-2\pi \kappa},
$$
and the unitarity of  evolution provides the probability of the transition to the degenerate subspace:
$$
P_{1\rar {\rm deg}} = 1-P_{1\rar 1} - P_{1\rar 4}.
$$

This solution  has a property that  distinguishes it from all previously solved models that had $t$-linear commuting operators. Namely, all such previously found transition probabilities depended only on the absolute values of the coupling constants. In the present model, the parameter combinations $|\gamma_{1,2}|^2$, and consequently the transition probability $P_{1\rar 4}$, which depends on them, generally depend on the phases of the original couplings $g_{k}$. 
Figure~\ref{p14-check-fig} shows results of our numerical simulations that confirm our analytical prediction (\ref{p14-fin}) and illustrate the dependence of  $P_{1\rar 4}$ on the phase $\phi$ of one of the coupling parameters. 

\section{Discussion}

We demonstrated that many MLZ models, which are generally considered unsolvable, have nontrivial $t$-quadratic  operators that satisfy integrability conditions (\ref{cond1})-(\ref{cond2}) with the original model's Hamiltonian. Here, we studied only one possible form of this operator, given by Eq.~(\ref{threeHp1}), and did not attempt to provide a general classification of the integrable families. Our goal was rather to  look at the properties of such integrable models.

Our results show that the integrability, in the sense of Eqs.~(\ref{cond1})-(\ref{cond2}), is generally not equivalent to the existence of the analytical solution. The main  effect of  integrability, within the MLZ theory, is  the existence of a continuous family of models with identical transition probabilities. This invariance does not follow from  some simple discrete symmetry of the Hamiltonian. Instead, this is an example of a nontrivial symmetry that is specific to explicitly time-dependent quantum systems.

Using the $t$-quadratic commuting operators, we were able to derive  a pair of new exactly solvable MLZ models. One model  was  the distortion of a previously known exactly solvable model, and the other one described the case with a degeneracy of two diabatic levels. A novel application of the $t$-quadratic operators, which we found, was the extension of the semiclassical Dykhne's approach to a multistate Hamiltonian.

\section*{Acknowledgements}
This work was supported by the U.S. Department of Energy, Office of Science, Basic Energy
Sciences, Materials Sciences and Engineering Division, Condensed Matter Theory Program.

\end{document}